\documentstyle[aps,epsf,prl]{revtex}
\newcommand{\beq}{\begin{equation}}
\newcommand{\eeq}{\end{equation}}
\newcommand{\beqas}{\begin{eqnarray*}}
\newcommand{\eeqas}{\end{eqnarray*}}
\newcommand{\beqar}{\begin{eqnarray}}
\newcommand{\eeqar}{\end{eqnarray}}
\newcommand{\req}[1]{(\ref{#1})}

\begin{document}
\twocolumn[\hsize\textwidth\columnwidth\hsize\csname
@twocolumnfalse\endcsname
\title{Expansion Around the Mean-Field Solution 
of the Bak-Sneppen Model}
\author{Matteo Marsili$^1$, Paolo De Los Rios$^2$, and Sergei Maslov$^3$}
\address{$^1$Institut de Physique Th\'eorique,
Universit\'e de Fribourg P\'erolles, Fribourg, CH-1700, 
Switzerland}
\address{$^2$Max-Planck-Institut f\"ur Physik Komplexer 
Systeme,
N\"othnitzer Str. 38, D-01187 Dresden, Germany}
\address{$^3$Department of Physics, Brookhaven National
Laboratory, Upton, New York 11973}
\date{\today}
\maketitle
\widetext
\begin{abstract}
We study a recently proposed equation for the 
avalanche distribution in the Bak-Sneppen model.
We demonstrate that this equation indirectly relates 
$\tau$, the 
exponent for the power law distribution of avalanche 
sizes, to $D$, the fractal dimension of an avalanche 
cluster.
We compute this relation numerically and approximate it
analytically up to the second order of expansion 
around the mean field exponents. 
Our results are consistent with Monte Carlo
simulations of Bak-Sneppen model in one and two 
dimensions.
\end{abstract}
]
\narrowtext

The Bak-Sneppen (BS) model \cite{bsmodel} has become 
one
of the paradigms of Self-Organized Criticality 
(SOC)\cite{btw}.
The rules of its dynamics are very simple:
the state of the model is completely defined by 
$L^d$ numbers $f_i$ arranged on a $d$-dimensional 
lattice of size $L$. 
At every time step the smallest of these numbers and 
its
$2d$  
nearest neighbors are replaced with new uncorrelated 
random
numbers, drawn from some distribution ${\cal P}(f)$.
This ``minimalistic'' dynamics results 
in a remarkably rich and interesting behavior.
In fact, there exists a whole class of models 
called {\it extremal} models \cite{PMB}, which 
evolve according to similar rules, and share
many similar features with the BS model. In all these
models the update happens only at the site carrying
the global minimum of some variable. The oldest, and 
perhaps  
the most widely known of these models is Invasion 
Percolation 
\cite{wilk}. The BS model, being the
simplest and the most analytically treatable extremal 
model, 
occupies the place of an ``Ising model'' in this class.

The self-organized critical nature of the BS model 
(as well as of other extremal models) is revealed in 
its ability to naturally evolve towards a 
stationary state where almost all the variables
$f_i$ are above a critical threshold $f_c$. 
The dynamics in the stationary state is characterized 
by scale-free bursts of activity or {\it avalanches}, 
which form a hierarchical structure 
\cite{bsmodel,backw}
of sub-avalanches within bigger avalanches.
The introduction of an auxiliary parameter
$f$ \cite{PMB} allows one to describe the system within
the paradigms of standard critical phenomena.
Indeed the distribution $P(s,f)$ of avalanche 
sizes $s$ close to $f_c$, 
has the same qualitative behavior
of the cluster distribution of 
percolation \cite{stauffer} above and below
the critical threshold $p_c$:
For $f<f_c$, $P(s,f)$ has a finite cut-off,
reminiscent of an undercritical system.
As $f\to f_c$ the cut-off diverges and a scale-free
distribution $P(s,f_c)\sim s^{-\tau}$ emerges.
In the overcritical regime $f>f_c$ 
there is a non-zero probability to start an infinite
avalanche, but all finite avalanches again 
have a finite cut-off. 
Scaling arguments\cite{PMB} allow one
to derive all critical exponents of a general 
extremal model in terms of only two independent ones,
say $\tau$ and $D$ -- the fractal dimension of the 
avalanche cluster. 

In order to compute these two remaining exponents
one has to resort to methods which go beyond scaling
arguments. Apart from the solution of the mean field 
case \cite{mf},  and a real space renormalization
group approach\cite{rsrg} for $d=1$, a 
systematic theory to compute the BS exponents
is still lacking. A promising step in this
direction was recently taken by one of us \cite{master} 
with the introduction of an {\it exact} equation for the 
avalanche distribution. 
Hereafter we will refer to this equation as
the Avalanche Hierarchy Equation (AHE). 
It was shown that inside the AHE is hidden
an infinite series of equations, relating
different moments of the avalanche size distribution.

In this letter we demonstrate that, as it was 
conjectured in 
\cite{master}, the AHE indirectly relates these two 
exponents, thus reducing the number of independent 
critical 
exponents in the BS model to just one. 
Contrary to simple rational relations based on scaling
arguments \cite{PMB}, this exponent 
relation is highly non-trivial. First we display the
numerical solution of the AHE.
Then we perform a perturbative ``$\epsilon$''-expansion 
around the mean field solution, up to the second order 
in 
$\epsilon$. 
The numerical solution of AHE is in agreement with
the MC simulations in $d=1,2$
\cite{PMB,grass}
and is well approximated
by the results of the $\epsilon$ expansion up to second
order.
They constitute a significant step forward towards
the full solution of the BS model. However, the unusual 
type of
the expansion around the mean-field exponents leaves 
open the
question of the upper critical dimension $d_c$ in the 
model.
It also does not answer the question about the 
geometrical,
fractal properties of the avalanche cluster. Instead, 
given
an avalanche fractal dimension $D$ it enables one to 
derive
the power $\tau$ of the avalanche distribution. 
Similarly,
in ordinary percolation the power of cluster 
distribution $\tau$
is related to the cluster's fractal dimension $D$ via a 
hyperscaling
relation $\tau=1+d/D$ \cite{stauffer}. 

Following ref. \cite{master}, let us consider the 
exponential distribution 
${\cal P}(f)=e^{-f}$, $f>0$. This simplifies the expressions 
without loss of generality \cite{notap}.
To define avalanches one records the {\it signal} of the 
model, i.e. the value of the global minimal
number $f_{min}(t)$ as a function of time $t$.
Then for every value of an auxiliary parameter $f$, an
$f$-avalanche of size (temporal duration) $s$
is defined as a sequence of $s-1$ successive events, 
when $f_{min}(t)<f$, confined between two events,  
when $f_{min}(t)>f$. In other words, the events, when
$f_{min}(t)>f$, divide the time axis into a series of 
avalanches, following one another. 
The AHE is an equation for the probability distribution 
$P(s,f)$ of $f$-avalanche sizes $s$, and it 
reads\cite{master}:
$\partial_f P(s,f) = 
\sum_{s_1=1}^{s-1} R^d(s_1) P(s_1,f)P(s-s_1,f)-R^d(s) P(s,f)$.
Here $R^d(s)$ is the average number of distinct sites
updated at least once during an avalanche of size $s$.
The equation describes how the sequence of 
avalanches changes when the value
of $f$ is raised by an infinitesimal amount $df$.
The first term on the RHS describes the
gain of avalanches of size $s$ due to merging
of two consecutive avalanches of size $s_1$ and 
$s-s_1$. Such merging occurs when the value of 
$f_{\min}(t)$, terminating 
the first avalanche, happens to be in the interval 
$[f,f+df]$.
The factor $R^d(s_1)$ comes from the fact that
when avalanches  merge, the active site starting
the second avalanche must be one of the sites updated in 
the
first avalanche. The bigger is the region covered by an 
avalanche, the better are its chances to merge with
the one directly following it.
The second term in the RHS is the ``loss term''
due to avalanches of size $s$ merging to form a larger 
avalanche.

To proceed further one needs to introduce the scaling 
ansatz 
$R^d(s) \sim s^{\mu}$ for the number of updated sites. 
This power law relation is a consequence 
of the {\it spatio-temporal} fractal structure of the 
avalanches
in the BS model \cite{maslov}.
The exponent $\mu(d)$, 
which is an independent ``input'' variable in the AHE,
depends on the dimensionality of the model and the
fractal structure of the avalanche. Physically, the 
exponent $\mu$ relates the volume of the spatial 
projection 
of an avalanche cluster to its
temporal duration $s$. 
If this spatial projection is a dense object with a 
fractal 
dimension equal to the dimension of space $d$, $\mu$ is 
given by $d/D$. In this expression $D$ is the fractal 
dimension of the 
avalanche\cite{PMB} defined through $s=R^D$. 
This is known to be true in $d=1$, where the 
connected nature 
of an avalanche cluster ensures the compactness (absence 
of holes)
of its projection. In $d=2$ the projection of the 
avalanche can have
holes, but still it  was
numerically found to be dense (i.e have a fractal 
dimension $d$) 
\cite{PMB}. It is clear
that, as the dimensionality of space is increased, the 
exponent
$\mu$ should approach $1$, since multiple updates of the same site 
become 
less and less likely and the volume of the projection
should be closer and closer to the total volume $(2d+1)s$ 
of the avalanche itself. The ``hyperscaling''
relation $\mu=d/D$ will be clearly violated for $d>d_c$, 
where one has $D=4$, $\mu=1$ \cite{PMB,expl_D}. 
{From} this it follows that $d_c\ge 4$ for the BS model.

\begin{figure}
\epsfxsize 12cm
\centerline{\epsfbox{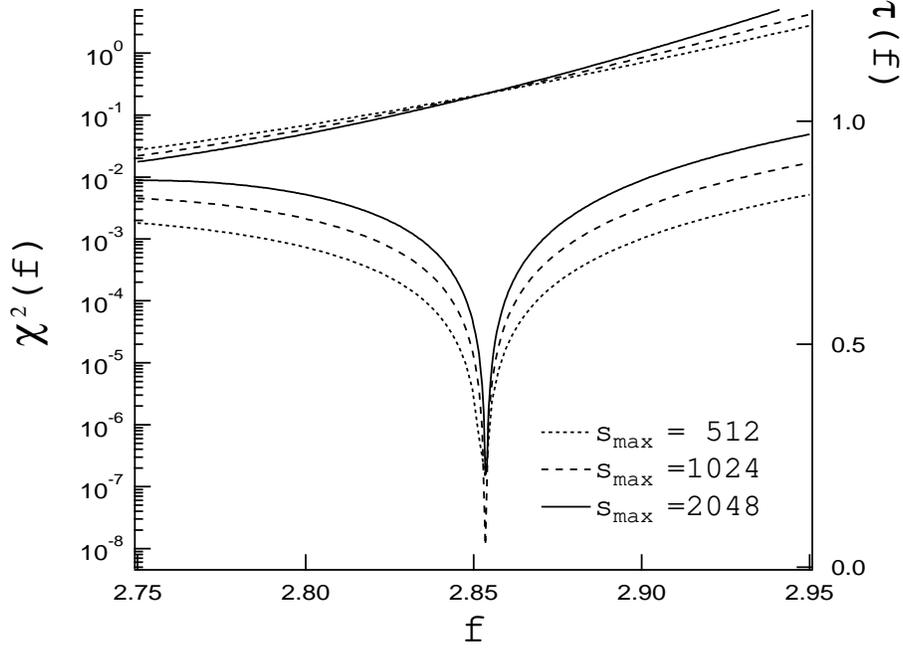}}
\caption{The plot of $\chi^2$ and $\tau$ vs $f$ for
$\mu=0.411$ and $s_{\max}=521,~1024$ and $2048$.}
\label{fig1}
\end{figure}

\begin{figure}
\epsfxsize 12cm
\centerline{\epsfbox{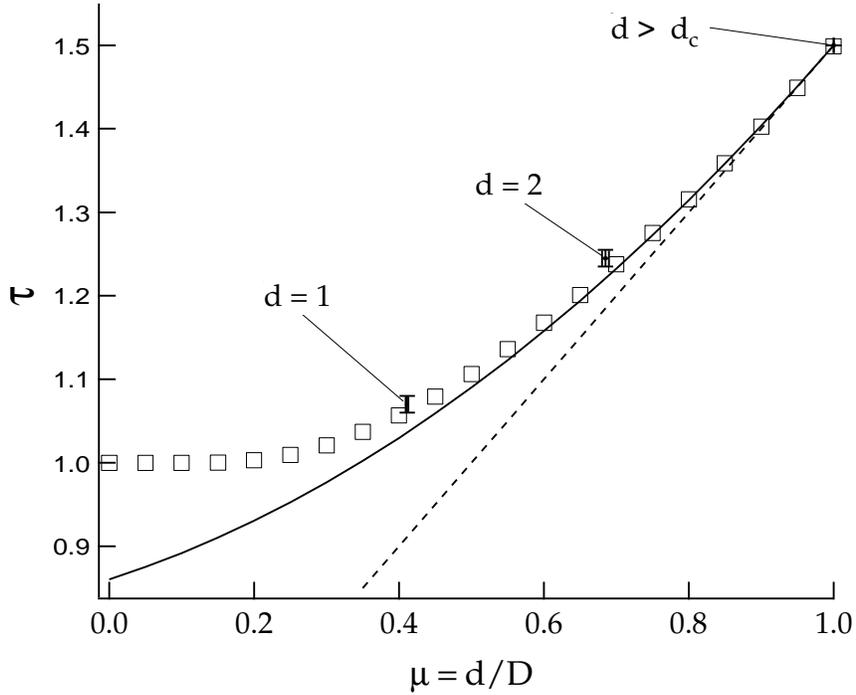}}
\caption{
The relation $\tau(\mu)$: numerical solution of Eq.(1) 
($\Box$, with $s_{\max}=1024$), 
expansion up to order $1-\mu$ (dashed line)
and $(1-\mu)^2$ (full line).
The results of the Monte Carlo numerical simulations 
in $d=1$, $2$ [3] 
and the mean field result [7] are also shown.}
\label{fig2}
\end{figure}

The introduction of the ``phenomenological'' 
exponent $\mu$ closes the AHE, which then reads
\beq
\partial_f P(s,f) =\sum_{s_1=1}^{s-1} 
s_1^\mu P(s_1,f)P(s-s_1,f)-s^\mu P(s,f).
\label{e.master0}
\eeq
The solution of equation (\ref{e.master0}) exhibits a power law 
behavior 
$P(s,f)\sim s^{-\tau}$ when $f$ is at its critical 
value $f_c$. Close to $f_c$ it takes a scaling form 
\beq
P(s,f)=s^{-\tau}F(s^{\sigma} \Delta f),
\label{sc.dist}
\eeq
where $\Delta f=f_c-f$. The exponents 
$\tau$, $\sigma$, and $\mu$ are related through
$\tau=1+\mu-\sigma$ \cite{PMB}. Perhaps a more
familiar form of this exponent relation involves
the correlation length exponent $\nu=1/\sigma D$.
The relation then becomes $\tau=1+ (d-1/\nu)/D$.
It has been conjectured \cite{master}
that equation \req{e.master0} also 
indirectly relates the exponents 
$\mu$ and $\tau$. In order to check this conjecture we 
numerically integrated Eq. \req{e.master0} forward in $f$ 
with the initial condition $P(s,0)=\delta_{s,1}$
for several values of $\mu$\cite{notai}. In order
to locate the critical point $f_c(\mu)$, a least square
fit of $\log P(s,f)$ vs $\log s$ was performed runtime
for each value of $f$. The value $\chi^2(f)$ of the sum 
of
the squared distances from the fit drops almost to zero 
in a 
very narrow region (see Fig. \ref{fig1}), 
which then allows for a very precise
estimate of $f_c(\mu)$ and $\tau(\mu)$. The results for 
the 
latter are shown in figure \ref{fig2} ($\Box$). 
These results are in a perfect agreement with Monte 
Carlo
estimates of $\tau$ in one and two dimensions \cite{PMB,grass}. 
In $d=1$, $\mu=1/D=0.411(2)$ and $\tau=1.07(1)$, while 
the results
of numerical integration of \req{e.master0} give 
$\tau(0.4)=1.058$. In $d=2$ the Monte Carlo results 
$\mu=2/D=0.685(5)$, $\tau=1.245(10)$ are also consistent
with our relation giving $\tau(0.7)=1.238$.
This confirms that Eq. \req{e.master0} indeed
contains a {\em novel} non-trivial relation between 
$\tau$ and $\mu$.

In order to address this relation analytically,
let us take the Laplace transform of 
Eq.\req{e.master0}~\cite{master}.
The AHE, with $p(\alpha 
,f)\equiv\sum_{s=1}^{\infty}P(s,f)
e^{-\alpha s}$, reads
\beq
\partial_f \ln[1-p(\alpha ,f)] =
\sum_{s=1}^{\infty}P(s,f)s^\mu e^{-\alpha s}.
\label{e.m2}
\eeq
$p(\alpha,f)$ has the scaling form given by:
\beq
p(\alpha,f)=1-\alpha^{\tau-1}
h\left(\Delta f/\alpha^{\sigma}\right).
\label{scalingform}
\eeq 
This scaling form follows from Eq.\req{sc.dist} and 
the 
scaling functions are related through
$h(x)=\int_0^{\infty}[F(0)-F(xy^{\sigma})e^{-y}]y^{-\tau
}dy$.

The scaling function $h(x)$ (as well as $F(x)$) is 
analytic at $x=0$.
Its large $|x|$ asymptotics is determined by the fact
that at any $\Delta f \neq 0$,  
$p(\alpha,f)$ is analytic in $\alpha$, since 
$p(\alpha,f)=1-P_{\infty}(f)+\langle s\rangle _f\alpha
+\langle s^2\rangle _f \alpha^2+\ldots$, and all 
the moments of $P(s,f)$ are finite except at the
critical point. 
Here $P_{\infty}(f)$ is the probability to start an 
infinite 
avalanche, and $P_{\infty}(f)=0$ for $f<f_c$.
Matching the expected behavior of $p(\alpha,f)$ to its 
scaling form one gets:
\beq
h(\pm |x|)=|x|^{(\mu-\sigma)/\sigma}\sum_{k=0}^\infty 
b_k^\pm |x|^{-k/\sigma}.
\label{asymp}
\eeq
In other words, $h(x)|x|^{(\sigma-\mu)/\sigma}$ for 
$|x|\gg 1$,
must be an analytic function of $|x|^{-1/\sigma}$. 
The coefficients $b_k^\pm$ for $k>0$
are related to the amplitudes of the diverging moments 
through
$\langle s^k\rangle_f = (- 1)^{k+1} 
b_{k}^\pm
|f_c-f|^{-(k+\sigma-\mu)/\sigma}$,
in the under-critical ($b_k^+$, $f<f_c$) and in the 
over-critical regime ($b_k^-$, $f>f_c$).
As we will see later, it is the condition that $h(x)$ 
has the desired asymptotic form \req{asymp},
which fixes the value of $\sigma$ for a given $\mu$. 

With the scaling ansatz \req{scalingform},
the LHS of Eq.\req{e.m2} becomes 
$-\alpha^{-\sigma}h'(x)/h(x)$, where
$x=\Delta f/\alpha^{\sigma}$.
The RHS needs some more work: using 
$s^\mu e^{-\alpha s}=-\partial_\alpha s^{\mu-1}
e^{-\alpha s}$ and the identity 
$s^{\mu-1}=\int_0^\infty 
t^{-\mu}e^{-st}dt/\Gamma(1-\mu)$,
one can express the RHS of Eq. \req{e.m2} in terms of 
an integral involving $p(\alpha,f)$.
Matching the powers of $\alpha$ in the resulting
equation for $h(x)$ one gets once more the 
well known exponent relation:
\beq 
\tau=1+\mu -\sigma.
\label{tau}
\eeq
After eliminating $\tau$, equation (\ref{e.m2}) finally reads:
\beq
h'(x)=\frac{h(x)}{\Gamma(1-\mu)}
\int_0^1 
dz\frac{xzh'(xz)-\frac{\mu-\sigma}{\sigma}h(xz)}
{(1-z^{1/\sigma})^{\mu}}
\label{eqh}
\eeq
We were not able to solve equation \req{eqh}
and find the exact relation $\sigma(\mu)$. 
However, we can explicitly  solve AHE for $\mu=1$.
This corresponds to 
the mean field version of the BS model, which has been 
studied in detail \cite{mf}. We will rederive their 
results using our approach. Our strategy will then be 
to perform a systematic $\epsilon$ -- expansion
around the mean-field solution, where $\epsilon=1-\mu$. 
This clearly
differs from the standard $\varepsilon=d-d_c$ expansion 
(note that the upper critical dimension $d_c$ for the 
BS model is still an open issue), since the 
dimensionality
of the system does not enter directly into our 
discussion.

As $\mu\to 1$, the integral in Eq. \req{eqh} 
diverges for $z \simeq  1$, but so does $\Gamma(1-\mu)$.
This implies that the factor $[\sigma \Gamma(1-\mu)(1-
z^{1/\sigma})^\mu]^{-1}$ behaves as a $\delta(z-1)$ 
function. For $\mu=1$ Eq. \req{eqh} 
reduces to
\beq 
h'(x)=\sigma x h(x) h'(x)-(1-\sigma )h^2(x).
\label{eq mean field}
\eeq
Its solution reads $h(x)[1-xh(x)]^{\sigma-1}=a_0$, 
with $a_0$ an integration constant. Eq. \req{asymp} 
implies a 
large $x$ behavior $h(x)\simeq 
x^{-1}(C+Dx^{-1/\sigma}+\ldots)$,
compatible with the solution of Eq.(\ref{eq mean field}) only if $C=1$ and 
$\sigma=1-\sigma=1/2$.
For $a_0=1$ one recovers a mean field solution \cite{mf}:
\beq
h^{(0)}(x)=\frac{\sqrt{4+x^2}-x}{2}. 
\label{solh1}
\eeq
The above derivation demonstrates that the knowledge of 
the whole
scaling function $h(x)$ is necessary in order to find 
the exponent $\sigma$.

To proceed beyond the mean field case
we perform a $1-\mu\equiv\epsilon$ -- expansion
of Eq. \req{eqh} around the $\mu=1$ solution 
\req{solh1}.
We put $\sigma=\frac{1}{2}+c\epsilon$, where $c$ is to 
be determined
later. It is convenient to change variables to 
$z=h^{(0)}(x)$ and to set
$h(x)=z[1+\epsilon\phi(z)+O(\epsilon^2)]$. 
Keeping only terms linear in $\epsilon$ 
one gets an equation for $\phi$:
\[\frac{1}{2}z(1+z^2)\partial_z\phi=\phi-1-\psi(1/2)
+(1+2c)z^2+2\ln\frac{1+z^2}{2z},\]
where $\psi(x)$ is the logarithmic derivative of 
$\Gamma(x)$.
This equation has to be solved with the boundary 
condition
$\phi(z=1)=0$ (i.e. $h(0)=1$). After some algebra one 
gets:
\beq
\phi(z)=A\frac{1-z^2}{1+z^2}-2\ln\frac{1+z^2}{2}
+\frac{2+(2-4c)z^2}{1+z^2}\ln z
\label{eqphi}
\eeq
where $A=2+\psi(1/2)\cong 0.03649\ldots$.
The value of $c$ is set by the requirement that 
$\phi(z)$ must give rise to the desired asymptotic
behavior of $h(x)$. The singular behavior at $z\simeq 
1/x\to 0$
must be matched to the asymptotics of $h(x)$ for 
$x\to\infty$. To order $\epsilon$, Eq. \req{asymp}
requires that 
$h(x)=x^{-1-2\epsilon}f(x^{-2+4c\epsilon})$, 
where $f(y)$ is analytic at $y=0$ [to order 
$\epsilon^0$, $f(y)=(\sqrt{1+4y}-1)/2$]. Expanding this 
relation to order $\epsilon$ one finds that the
singular part of $\phi(z)$ 
must have {\em exactly} 
the form $2[1+2cz^2/(1+z^2)]\ln z$. The only 
value
of $c$ which matches this requirement to the last term 
in the RHS of Eq. \req{eqphi} is $c=0$.
Note that the whole asymptotic behavior, and not just 
its leading part, is necessary to determine $c$.

This concludes the first order of the expansion in 
$\epsilon$. We have found that in the first order in 
$\epsilon$ the critical exponent $\sigma$ did not 
change. The exponent relation \req{tau} then gives 
$\tau=3/2-\epsilon$. Finally, the analytic form of the
scaling function $h(x)$, containing all information 
about
the amplitudes of avalanche moments, is given by
\[
h(x)={\sqrt{4+x^2}-x \over 2}\left[1+{x^2 \over 
4}\right]^{-\epsilon}
\left[1+\frac{\epsilon\, 
A\,x}{\sqrt{4+x^2}}\right]+O(\epsilon^2).
\]
  
The extension of this procedure to higher
orders in $\epsilon$ is straightforward, 
even though it involves much heavier algebra. 
Skipping the details \cite{details}, 
up to second order in $\epsilon$ we find
\beqar
\sigma &=&\frac{1}{2}-\frac{4}{3}(\gamma+\ln 2 
-1)\epsilon^2+O(\epsilon^3)
\nonumber\\
&\simeq&0.5-0.3605 \epsilon^2+O(\epsilon^3);
\label{sigma}\\
\tau &\simeq&1.5-\epsilon+0.3605 
\epsilon^2+O(\epsilon^3).
\label{tau2}
\eeqar
Here $\gamma \simeq 0.5772$ is the Euler's constant.
The explicit expression for $h(x)$ at this order
is not particularly illuminating, so we refrain to
display it here.
As seen in Fig. \ref{fig2}, the expansion
up to the first two orders is 
in excellent agreement with numerical data down to 
$\mu\approx 0.6$ ($\epsilon \approx 0.4$). 

On the other side, Fig. \ref{fig2} seems to suggest a 
singular behavior of $\tau(\mu)$ as $\mu\to 0$.
The specialty of $\mu=0$ can be understood by observing 
that 
in this case $P(s,f)$ does not obey scaling.
Indeed, $\mu=0$ corresponds to a trivial model with 
only one constantly updated site, 
which can be considered as a $0$-dimensional lattice. 
The probability of $f$-avalanches of size
$s$ is trivially derived from the probability 
$P(f_{\min}(t)<f)=1-e^{-f}$ that the signal
is below $f$: $P(s,f)=e^{-f}(1-e^{-f})^{s-1}$.
This is indeed the solution of Eq. \req{e.master0} with
$s^\mu =1$. There is no phase transition 
(numerically we found $f_c(\mu)\sim 1/\mu \to \infty$ 
as $\mu \to 0$) and the avalanche distribution always 
has 
an exponential cutoff. This suggests that
$d=0$ can be interpreted as the lower critical
dimension for the BS model (note that $P(s,f)$ in 
the $d=0$ BS model is very similar to the cluster size 
distribution in the $d=1$ percolation\cite{stauffer}). 

In conclusion, we have shown that the avalanche
hierarchy equation introduced by one of us in \cite{master}
yields a new relation between exponents in 
the Bak-Sneppen model, thus reducing the number of
independent exponents to just one. This relation
expresses $\tau$, the power law exponent in
the avalanche probability distribution, in terms of $D$, 
the mass dimension of an avalanche cluster. 
We were able to perform a systematic expansion of this 
relation around the mean field exponents, carried to the 
second order in this work. The success of this
approach suggests that a complete 
$\varepsilon =d-d_c$ expansion for the BS model could 
be possible. The accomplishment of this task, however, 
calls for a systematic study of the BS model in high
dimensions and for identification of the 
upper critical dimensionality $d_c$.

The authors thank the I.S.I. foundation in Torino
where this work was begun. S.M. and M.M. thank the 
Institut 
de Physique Th\'eorique, at the University of Fribourg, 
and the Max Planck Institute in Dresden, respectively, 
for the hospitality.
P.D.L.R. thanks A. Valleriani for many useful discussions and for a careful
reading of the manuscript.

\vfill\eject
\onecolumn

\begin{references}
\bibitem{notap}
The BS model is ``insensitive''
to the shape of random number distribution
because only the {\it order} among
variables $f_i$ matters in the dynamics. With the 
rescaling map 
$f(f')= -\ln \int_{f'}^{\infty} P(f') df'$,
the BS dynamics for a general $P(f')$ is
turned into that with the exponential 
distribution.
\bibitem{expl_D} Indeed a cluster in the  BS
model above the upper critical dimension has the same
fractal dimension $D=4$ as the cluster in the mean-field ordinary
or directed percolation \cite{stauffer}, as well as any other
mean-field branching process.
\end{references}
\end{document}